\documentclass[12pt]{article}
\usepackage{epsfig,textpos,url }
\usepackage{lineno}

\newcommand\pt{\partial}
\begin{document}
\title{Wien-Filter Hamiltonian and Transfer Matrix}
\author{Volker Ziemann, Jefferson Lab}
\date{August 5, 2026}
\maketitle
\begin{abstract}\noindent
  We derive the Hamiltonian and the transfer matrix of a Wien filter
  with vertical magnetic field from first principles.
\end{abstract}
%
%
%
\section{Introduction}
Wien filters are essential for operating CEBAF with spin-polarized beams,
but they also affect the normal first-order beam optics, even under ideal
circumstances. We therefore derive the Hamiltonian and the six-dimensional
transfer matrix for an ideal Wien filter. By ideal we mean that the magnetic
and electric fields are considered hard edged and that no misalignment of
electrodes or any other kind of imperfection are taken into account. Transfer
matrices in four~\cite{ENGE} and five~\cite{HURD} dimensions were derived 
earlier. We complement these works by considering the full six-dimensional 
phase space using the variables $x$, $x'$, $y$, $y'$, $z$, and $\delta=\Delta p/p$. 
This might be useful to include the effect of Wien filters in conventional
beam optics codes such as the descendants of Karl Brown's ancient TRANSPORT 
code~\cite{TP}.
\section{Hamiltonian}
We base our analysis on the formalism outlined in Section~2 in~\cite{VZAPB} and
start from the general Hamiltonian
{\small
\begin{eqnarray}\label{eq:ham1}
  &&H(x,x',y,y',z,\delta;s)=-\left(1+\frac{x}{\rho}\right)\frac{eA_s}{p_o}+\delta\\
  &&\quad-\left(1+\frac{x}{\rho}\right)
      \sqrt{\left(\beta_0\delta+\frac{1}{\beta_0}+\frac{e\Phi}{cp_0}\right)^2
      -\left(x'-\frac{eA_x}{p_0}\right)^2-\left(y'-\frac{eA_y}{p_0}\right)^2
      -\frac{1}{\beta_0^2\gamma_0^2}}\ ,\nonumber
\end{eqnarray}}%
where we tacitly assume the paraxial approximation with small angles $x'$ and $y'$ and
where $\Phi$ is the electric potential and $\vec A$ the magnetic potential~\cite{VZAPB}.
In a Wien filter the electric and magnetic fields compensate each other such that
the trajectory is straight, which implies $1/\rho=0$. The magnetic and electric
potential, in turn, are given by
\begin{equation}
  \Phi= E x \qquad\mathrm{and}\qquad A_s=-B_0x\ .
\end{equation}
The other two components of the magnetic  potential $A_x$, and $A_y$ are zero,
$B_0$ is the vertical component of the magnetic field, and $E=V/d$ is the
horizontal electric field strength produced by a voltage $V$ between
electrodes that are separated by a distance $d$. Inserting these quantities
in Equation~\ref{eq:ham1} leads us to
{\small
\begin{equation}
  H(x,x',y,y',z,\delta;s)=\frac{eB_0x}{p_0}+\delta -
  \sqrt{\left(\beta_0\delta+\frac{1}{\beta_0}+\frac{eEx}{cp_0}\right)^2
    -x^{\prime 2}-y^{\prime 2}-\frac{1}{\beta_0^2\gamma_0^2}}\ .
\end{equation} }
Expanding the root to second order in the dynamical variables $x,x',y,y',z$, and
$\delta$ brings us to
\begin{eqnarray}
  H&=&\left[\frac{eB_0x}{p_0}+\frac{eEx}{\beta_0cp_0}\right]-1
     +\frac{\delta^2}{2}\left(1-\beta_0^2\right)
     -\frac{1}{2}\left(\frac{eEx}{cp_0}\right)^2
     +\frac{1}{2}\left(\frac{eEx}{\beta_0cp_0}\right)^2\nonumber\\
  &&\qquad +\beta_0\frac{eEx\delta}{cp_0}-\frac{eEx\delta}{\beta_0cp_0}
     +\frac{1}{2}x^{\prime 2}+\frac{1}{2}y^{\prime 2} \nonumber\\
   &=& \left[\frac{eB_0x}{p_0}+\frac{eEx}{\beta_0cp_0}\right]-1
       +\frac{\delta^2}{2\gamma_0^2}
       -\frac{1}{2}\left(\frac{eEx}{\beta_0cp_0}\right)^2\left(\beta_0^2-1\right)\\
   &&\qquad +\frac{eEx\delta}{\beta_0cp_0}\left(\beta_0^2-1\right)
      +\frac{1}{2}x^{\prime 2}+\frac{1}{2}y^{\prime 2}\nonumber\\
   &=&\left[\frac{eB_0x}{p_0}+\frac{eEx}{\beta_0cp_0}\right]-1
       +\frac{\delta^2}{2\gamma_0^2} +\frac{1}{2}x^{\prime 2}+\frac{1}{2}y^{\prime 2}\nonumber\\
  &&\qquad+\frac{1}{2\gamma_0^2}\left(\frac{eEx}{\beta_0cp_0}\right)^2
      -\frac{1}{\gamma_0^2} \frac{eEx\delta}{\beta_0cp_0}\ . \nonumber
\end{eqnarray}
First we note that the constant $-1$ does not affect the equations of motion and that the
expression in the square bracket vanishes, if the Wien condition $B_0=-E/\beta_0c$
is fulfilled, which we assume to be true in the following sections. Thus we arrive at
\begin{equation}
  H(x,x',y,y',z,\delta;s)= \frac{\delta^2}{2\gamma_0^2}
  +\frac{1}{2}x^{\prime 2}+\frac{1}{2}y^{\prime 2}
  +\frac{1}{2}\left(\frac{eEx}{\gamma_0\beta_0cp_0}\right)^2
  -\frac{1}{\gamma_0} \frac{eEx\delta}{\gamma_0\beta_0cp_0}\ .
\end{equation}
The three first expressions are the same that describe a drift space, the
term proportional to $x^2$ describes horizontal focusing and the last
term, proportional to $x\delta$ describes the energy-dispersive properties
of the Wien filter. We note that $-eE/\gamma_0\beta_0cp_0=B_0/(\gamma_0p_0/e)=(1/\gamma_0)B_0/(B\rho)=1/R$
has the units an inverse length $R$, where we use $p/e=(B\rho)$ to parametrize
the energy of the beam. This enables us to simplify the Hamiltonian further
\begin{equation}
  H(x,x',y,y',z,\delta;s)= \frac{\delta^2}{2\gamma_0^2}
  +\frac{1}{2}x^{\prime 2}+\frac{1}{2}y^{\prime 2}
  +\frac{x^2}{2R^2} + \frac{x\delta}{\gamma_0R}
\end{equation}
with $R=-\gamma_0\beta_0cp_0/eE=\gamma_0(B\rho)/B_0$. This Hamiltonian is the starting point
to derive the equations of motion.
\section{Equations of motion}
In the horizontal plane, we find the equations
\begin{equation}\label{eq:emx}
  \frac{dx'}{ds}=-\frac{\pt H}{\pt x} = -\frac{x}{R^2}-\frac{\delta}{\gamma_0R}
  \qquad\mathrm{and}\qquad
  \frac{dx}{ds}=\frac{\pt H}{\pt x'}=x'\ ,
\end{equation}
which have the solution
\begin{eqnarray}\label{eq:emsx}
  x(s)&=&x_1\cos(s/R)+x'_1R\sin(s/R) -(R/\gamma_0)\left(1-\cos(s/R)\right)\delta\nonumber\\
  x'(s)&=&-(x_1/R)\sin(s/R)+x'_1\cos(s/R)-(\delta/\gamma_0)\sin(s/R)\ .
\end{eqnarray}
Here and in the following equation quantities with subscript $1$ denote values
at the entrance of the Wien filter. In the vertical plane, we have
\begin{equation}\label{eq:emy}
  \frac{dy'}{ds}=-\frac{\pt H}{\pt y}=0
  \qquad\mathrm{and}\qquad
  \frac{dy}{ds}=\frac{\pt H}{\pt y'}=y'
\end{equation}
with the solution
\begin{equation}\label{eq:emsy}
  y'(s)=y'_1\qquad\mathrm{and}\qquad y(s)=y_1+y'_1s \ .
\end{equation}
Finally, for the longitudinal plane, we find
\begin{equation}\label{eq:emz}
  \frac{d\delta'}{ds}=-\frac{\pt H}{\pt z}=0
  \qquad\mathrm{and}\qquad
  \frac{dz}{ds}=\frac{\pt H}{\pt\delta}=\frac{\delta}{\gamma_0^2} + \frac{x}{\gamma_0R}\ .
\end{equation}
The first equation has the solution $\delta(s)=\delta_1$. Inserting $x$ from Equation~\ref{eq:emsx} 
and integrating the equation for $dz/ds$ with respect to $s$, we obtain the solution
\begin{eqnarray}\label{eq:emsz}
  z(s)&=& \frac{\delta_1}{\gamma_0^2}s+\frac{x_1}{\gamma_0}\sin(s/R)+R\frac{x'_1}{\gamma_0}\cos(s/R)\nonumber\\
 &&\qquad       -\frac{\delta_1}{\gamma_0^2}\left(s-R\sin(s/R)\right) \\
 &=& \frac{x_1}{\gamma_0}\sin(s/R)+R\frac{x'_1}{\gamma_0}\cos(s/R) +\frac{R}{\gamma_0^2}\sin(s/R)\delta_1\ .
\nonumber
\end{eqnarray}
Here we need to keep in mind that $z_2-z_1=z(L)-z(0)$.
We now use equations~\ref{eq:emsx}, \ref{eq:emsy} and~\ref{eq:emsz} to construct the transfer matrix
that  maps the phase-space coordinates at the entrance (with subscript~$1$) to those at the
exit of the Wien filter (with subscript~$2$).
\section{Transfer matrix}
We assume that the Wien filter has the length $L$, and with the abbreviation $\phi=L/R$ we find
\begin{eqnarray}
  x_2&=&\cos(\phi) x_1+ R\sin(\phi) x'_1-\frac{R\delta_1}{\gamma_0}\left(1-\cos(\phi)\right) \nonumber\\
  x'_2&=&-\frac{1}{R}\sin(\phi) x_1 +\cos(\phi) x_2-\frac{\delta_1}{\gamma_0}\sin(\phi)\nonumber\\
  y_2&=&y_1+y'_1L\nonumber\\
  y'_2&=&y'_1 \\
  z_2&=& z_1+\frac{x_1}{\gamma_0}\sin(\phi)+\frac{x'_1R}{\gamma_0}\left(1-\cos(\phi)\right)
        +\frac{R\delta_1}{\gamma_0^2}\sin(\phi) \nonumber\\
  \delta_2&=&\delta_1\ .\nonumber
\end{eqnarray}
These equations are easily assembled into a (symplectic) matrix equation
\begin{equation}\label{eq:TM}
  \left(\begin{array}{c} x_2 \\ x'_2\\ y_2\\ y'_2 \\ z_2 \\ \delta_2\end{array}\right)
  = \left(\begin{array}{cccccc}
     \cos\phi & R\sin\phi & 0 & 0 & 0 & -\frac{R}{\gamma_0}(1-\cos\phi)\\
     -\frac{1}{R}\sin\phi & \cos\phi & 0 & 0 & 0 & -\frac{1}{\gamma_0}\sin\phi\\
            0 & 0 & 1 & L & 0 & 0\\
            0 & 0 & 0 & 1 & 0 & 0\\
            \frac{1}{\gamma_0}\sin\phi &\frac{R}{\gamma_0}(1-\cos\phi) & 0 & 0 & 1 & \frac{R}{\gamma_0^2}\sin\phi\\
            0 & 0 & 0 & 0 & 0 & 1\\
    \end{array}\right)        
  \left(\begin{array}{c} x_1 \\ x'_1\\ y_1\\ y'_1 \\ z_1 \\ \delta_1\end{array}\right)
\end{equation}
with $R=\gamma_0(B\rho)/B_0$. We point out that $R$ is not the bending radius caused by the 
magnetic field; it differs by a factor $\gamma_0$. Otherwise the transverse part
of the matrix appears to be very similar to that of a horizontal sector dipole. We point out
that those elements calculated in~\cite{HURD} agree with Equation~\ref{eq:TM}.
\par
It is instructive to relate the beam optics to the effect the Wien filter has on the spin.
The spin-rotation angle can be calculated from the Thomas-BMT equations~\cite{TBMT} 
in the rest frame of the electrons. Transforming back to the laboratory frame leads us
to~\cite{MAMI} 
\begin{equation}
  \Theta=\frac{eB_0L}{\gamma_0^2\beta_0mc} = \frac{eB_0L}{\gamma_0p_0} = \frac{B_0L}{\gamma_0(B\rho)}
  =\frac{L}{R}=\phi
\end{equation}
such that the transfer matrix is already suitable parametrized to characterize the effect of
spin-rotation on the beam optics.
\par
For 200\,kV electrons in the CEBAF injector, $\gamma_0\approx 1.39$ and
$(B\rho)=1.65\times 10^{-3}\,$Tm. The Wien filter is about $L=0.43\,$m long,
such that a $\phi=90^o$ or $\pi/2$ spin-rotation requires $R=L/(\pi/2)\approx0.27\,$m and
that translates into a magnetic field of $B_0=\gamma_0(B\rho)(\pi/2)/L=8.4\times10^{-3}\,$T.
Moreover, we can determine the magnitude of the betatron focusing by calculating the
approximate focal length of the Wien filter from the $R_{12}$ matrix element via
$1/f\approx L/R^2$ which gives us $f\approx 0.17\,$m. We emphasize that this
approximate formula is used outside its range of validity, because $L$ is bigger
than $R$ and thus the magnet is by no means ``short''. In June 2025~\cite{MAX}, we
measured a dominant focal length of $f_y=0.78\,$m for a $40^o$ spin rotation of the
vertical Wien filter in the CEBAF injector; also a skew component and some (weaker)
horizontal focusing was present at the time. Using the equations above we
estimate a focal length  of $f=0.88\,$m for these settings, which is close enough to
give us confidence that we capture the essential physics of the Wien filter.
\par
A potentially important point to note is that flipping the sign of the polarization by
reversing the polarity of the fields in the Wien filter also reverses the sign of $R$. The
top-left $4\times4$ part of the transfer matrix is unaffected, leaving the beta functions
the same. The dispersive contribution in the sixth column and fifth row, on the 
other hand, will change sign. The magnitude of the dispersion with spin-rotation
angle of $\phi=90^o$ is $R_{16}=R/\gamma_0\approx 0.2\,$m, which is likely non-negligible.
Moreover, the $R_{51}$ transfer-matrix element couples the horizontal orbit $x_1$ into
the longitudinal plane. At $\phi=90^o$ a change of $\Delta x_1=1\,$mm causes a change in
the longitudinal position $\Delta z_2=\Delta x_1/\gamma_0\approx 0.72\,$mm. With an RF
wavelength $\lambda=0.2\,$m this translates into a phase change of
$\psi_{RF}=360^o\times 0.72\times 10^{-3}/0.2\approx 1.3^o$, which is also quite substantial.
The root of the problem is that a $90^o$ spin rotation requires a magnetic field that
corresponds to a magnet excited to generate a rather large deflection angle of
$\gamma_0\times 90^o$, unless compensated by the electric field. 
Reversing the spin direction thus affects the beam transport
in a systematic way, which probably has to be taken into account in high-precision 
experiments like MOLLER~\cite{MOLLER}. 
\section{Conclusion}
We calculate the Hamiltonian and the transfer matrix for a Wien filter and found that the
resulting matrix is already parametrized by the spin-rotation angle. Reversing the 
angle will leave the beta functions unaffected, but will flip the sign of the dispersion
created by the Wien filter. Also the arrival time in cavities $\tau=z/c$ will be affected
via the matrix elements $R_{51}$ and $R_{52}$, unless the beam is well-centered with $x_1$
and $x'_1$ being very close to zero.
\par
In this note, we considered a Wien filter with a vertical magnetic field. The corresponding
matrix for a horizontal magnetic field can can be found by sandwiching  the transfer matrix
from Equation~\ref{eq:TM} between two $90^o$ coordinate rotations, one with positive and
the second with negative rotation angle in much the same way the transfer matrix of a
vertical dipole is constructed from that of a horizontal dipole in most beam optics codes.
\par
We emphasize that we do not take fringe fields into account and refer to numerical methods
such a GPT~\cite{GPT} or to reference~\cite{THIRD}.
\section*{Acknowledgements}
This material is based upon work supported by the U.S. Department of Energy, Office of Science,
Office of Nuclear Physics under Contract No. 89243126CSC000213.
%
%
\bibliographystyle{plain}

\end{document}